\PassOptionsToPackage{normalem}{ulem}
\documentclass[%
 reprint,
superscriptaddress,
 amsmath,amssymb,
 aps,
]{revtex4-2}

\usepackage{graphicx}% Include figure files
\usepackage{color} 
\usepackage{dcolumn}% Align table columns on decimal point
\usepackage{bm}% bold math
\usepackage[utf8]{inputenc}
\usepackage{amsmath}

\begin{document}

\preprint{APS/123-QED}

%TC:ignore
    
%\title{Melting and order in a 2D non-equilibrium dynamical model}% 
%\title{Crystallization and reentrant melting in a 2D non-equilibrium system}
\title{Active diffusing crystals in a 2D non-equilibrium system}

\author{Ashley Z. Guo}
\thanks{These two authors contributed equally}
\affiliation{%
Department of Chemical and Biochemical Engineering, Rutgers University–New Brunswick, Piscataway, New Jersey 08854, USA
}%

\author{Sam Wilken}
\thanks{These two authors contributed equally}
\affiliation{%
Department Chemie, Johannes Gutenberg University - Mainz, Mainz 55122, Germany
}%

\author{Dov Levine}
\affiliation{
Department of Physics, Technion-IIT, 32000 Haifa, Israel
}%
\author{Paul M. Chaikin}
\affiliation{%
Center for Soft Matter Research, Department of Physics, New York University, New York 10003, USA
}%

\date{\today}

\begin{abstract}
We investigate a 2D dynamical absorbing state model of monodisperse disks, in which rich phase behavior arises from interactions consisting solely of repulsive displacements between overlapping particles. 
The phase diagram reveals several unconventional features, including a disordered and static {\it absorbing} configuration, where no particles overlap, separated by a second-order phase transition to a \textcolor{black}{continuously evolving {\it active}} hexagonal crystal with collective ring diffusion, which in turn undergoes a first-order phase transition to an active isotropic liquid. The only driving parameter is $\epsilon$, the maximum size of the random repulsive kicks.  Small $\epsilon$ facilitates self-organization into an ordered state, but large $\epsilon$ prevents this organization from occurring. This is very different from typical order-disorder transitions, where there are two competing influences, energy and entropy, that drive the transition.
\end{abstract}

\maketitle

The dimensionality of space plays a crucial role in the melting transition from crystal to disorder. 
In conventional 3D crystals, equilibrium melting is a first-order, discontinuous transition that results from the competition between potential-driven order and temperature-driven disorder~\cite{born1996dynamical,lowen1994melting,de2023melting}. 
In contrast, 2D equilibrium melting transitions are often continuous and arise from a proliferation of topological defects and may proceed via an intermediary hexatic phase~\cite{kosterlitz1973ordering,halperin1978theory,young1979melting,nelson2002defects,zahn1999two,kelleher2017phase,stan1989two}. 
The melting transition can become even more complex out of equilibrium~\cite{Henkel2008,digregorio2018full,galliano2023two,wang2016flow}.
%, and non-equilibrium dynamical models have described many different categories of intermediary states, which also lead to different types of spatial ordering. 

\begin{figure*}%[bt!]
    \centering
    \includegraphics[width=1\textwidth]{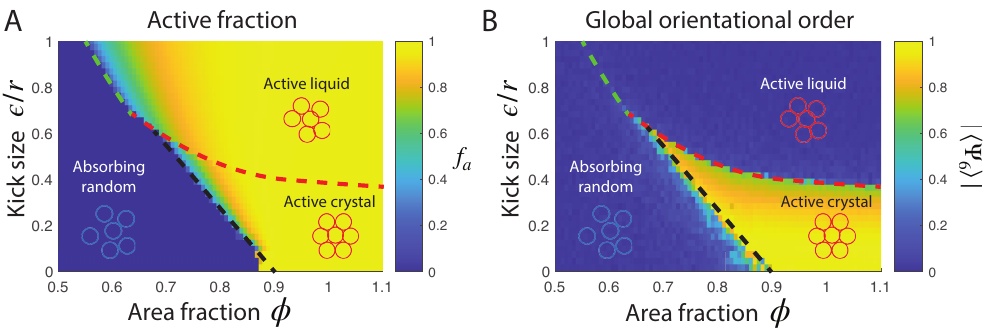}
    \caption{
    Monodisperse 2D Biased Random Organization (BRO) phase diagram. 
    As a function of the two control parameters, the repulsive kick size $\epsilon/r$ and area fraction $\phi$, BRO configurations evolve from random initial conditions to three steady-state phases: 1) absorbing random, 2) active liquid, and 3) active crystal. 
    These phases are characterized by the fraction of active particles $f_a$ ({\bf A}) and the global orientational order $|\langle \Psi_6 \rangle |$ ({\bf B}).
    Global translational order, as measured by the height of the first Bragg peak, looks qualitatively similar to the global orientational order in {\bf B}.
    All absorbing states, where particles avoid overlaps, are random. 
    Active states show both disordered (liquid) and ordered (crystalline) structural features, separated by a first-order phase boundary (red dashed line). 
    The boundary between absorbing and active liquid phases in monodisperse BRO exhibits Manna critical exponents and matches that of bi-disperse 2D BRO (green dashed line). 
    The transition from absorbing to active crystal phases also displays Manna exponents, with the densest absorbing state corresponding to the maximum circle packing $\phi(\epsilon \rightarrow 0) = \pi/(2\sqrt{3})$. 
    All simulations contain $N = 1600$ particles and are run until a steady state value of $f_a$ and $|\langle \Psi_6 \rangle |$ is reached.
    %{\bf C} A representative configuration of the active crystal (red circles) plotted alongside particle trajectories. Active crystals are defect-free, yet trajectories reveal rearrangements that result in long-timescale transport.
    %Phase diagram for the 2D Biased Random Organization model as a function of area fraction $\phi$ and repulsive kicksize $\epsilon/r$, where $r$ is particle radius. The system exhibits one absorbing phase and two different active phases, one disordered and one crystalline. Voronoi diagrams constructed from the particle centers are shown for the two active phases. To highlight crystalline regions, grain boundaries, and defects in the system, six-sided Voronoi cells are shown in white, others are colored. 
    %The absorbing-active transition analyzed in Figure \ref{fig3} follows the dashed orange line, and the active-active transition shown in Figure \ref{fig3} follows the dashed grey line.
    }
    \label{fig1}
\end{figure*}

Here, we study non-equilibrium melting in two dimensions, in the context of a continuous absorbing-state model~\cite{lubeck2004universal}. 
%\sout {which quantitatively reproduces the structural and dynamical features associated with dynamical phase transitions in periodically-sheared colloidal suspensions~\cite{Wilken2020,Pine2005,Corte2008}}.
The model we study,  Biased Random Organization (BRO), has a dynamical phase transition from absorbing to active in every dimension~\cite{Wilken2023}, and has been shown to have qualities familiar from equilibrium continuous phase transitions~\cite{Henkel2008}. 
However, the densest critical configurations %obtained as $\epsilon \rightarrow 0$ and $\phi \rightarrow \phi_c$ 
vary qualitatively across dimensions: in 1D and 2D, the particles crystallize, while in 3D through 5D, the transition state appears to be Random Close Packed~\cite{Wilken2020,wilken2021random,Wilken2023}. 
%and whose densest critical point coincides with Random Close Packing three, four, and five dimensions~\cite{Wilken2020RCP,Wilken2023}.
%is a variant of the Random Organization model originally developed to study sheared colloidal suspensions\cite{Pine2005, Corte2008}. 

% Need the "In this letter" paragraph here
% 1) Re-entrant disorder
% 2) one-step melting separates disordered and ordered Manna dynamical phase transitions
% 3) Dynamical active crystal phase collective 'ring' diffusive transport (violates Lindemann criterion)

In addition to the nature of the critical configurations, the 2D phase diagram itself shows remarkable features. In particular, we observe re-entrant disorder in absorbing and active phases, separated by a new active crystalline phase. 
This active crystalline phase exhibits diffusive transport not through vacancy or interstitial defects, but through a long-proposed~\cite{zener1950ring} yet rarely observed~\cite{freysoldt2014first} collective ring diffusion mechanism, violating the Lindemann criterion~\cite{lindemann1910ueber}. 
As activity increases, surprisingly, and unlike equilibrium melting, we find a simultaneous loss of translational and orientational order when crossing the active crystal-active liquid boundary via a discontinuous phase transition.
Our investigation suggests a potential paradigm shift in understanding non-equilibrium phase transitions, where a single activation source can both order and disorder, in contrast to the competition between temperature and potential interactions for ordering in equilibrium.

In 2D BRO, $N$ identical discs of radius $r$ are initially distributed randomly in the plane with an area fraction $\phi$. 
Overlapping particles are moved apart by a displacement that is random in magnitude, up to a maximum kick size $\epsilon$ \footnote{In the version we study here, the displacements of a pair of overlapping particles are along the line joining their centers.  If a particle overlaps with more than one other particle, only the closest overlap is considered, and both particles are displaced away from each other along the line joining their centers.}. 
These dynamics are then repeated until one of two things develops: (1) a quiescent absorbing state, where no overlaps remain, or (2) an active steady state, where the average fraction of overlapping, active particles reaches a constant value. 

If we fix  $\epsilon$ and vary $\phi$, we find that there is an $\epsilon$-specific critical density $\phi_c$ below which an absorbing state is always found, and above which the system is in a dynamic and continually evolving active state. 

%moved above discussion of 2D BRO
%In BRO, this dynamical phase transition from absorbing to active takes place in every dimension, and has been shown to have qualities familiar from equilibrium continuous phase transitions\cite{Henkel2008}. However, the critical configurations obtained as $\epsilon \rightarrow 0$ and $\phi \rightarrow \phi_c$ are very different across different dimensions: In 1D and 2D, the particles crystallize, while in 3D through 5D, the transition state appears to be Random Close Packed\cite{Wilken2020,Wilken2023}. 

%In addition to the different structural organization of the configurations in different dimensions, the phase diagrams themselves are markedly different.  
The 2D phase diagram is much richer than those in other dimensions.
%, and we focus on this phase behavior in this paper. 
While there is a single phase transition from an absorbing phase to an active phase in other dimensions, we find two phase transitions in 2D: the familiar absorbing-active transition, and another between distinct crystalline and liquid-like active phases. 
Thus, there are two order parameters needed to describe the system, one dynamical (the active fraction $f_a$) and the other structural (the hexatic parameter $\Psi_6$).
%Unlike in 3D, where BRO exhibits a highest density critical point corresponding to a disordered, randomly close-packed, jammed state with a single phase boundary, 2D BRO does not produce an equivalent random close-packed state and instead exhibits multiple phase boundaries. 

%\textcolor{green}{\sout{ It has already been established that the dynamical transition at low kick size in 2D is accompanied by the formation of a hexagonal crystal\cite{Wilken2023}. We show here that this crystal phase is kept intact well into the active region, and is only destroyed when either the kick size, or more surprisingly, the density is increased. In fact, we find that the crystal survives to an arbitrarily high density in the $\epsilon \rightarrow 0$ limit. }}

Figure \ref{fig1} presents the phase diagram, where we see three phases: a low-density disordered absorbing phase, a small $\epsilon$ but high $\phi$ active crystalline phase, and a high $\phi$ and high $\epsilon$ active disordered phase.
The existence of a hexagonal crystal was reported in \cite{Wilken2023}, and further studied in \cite{galliano2023two} in the context of a similar model, where it was found that the ordered phase exhibited long-range order in contrast to the Mermin-Wagner theorem applicable in thermal equilibrium.
%but different model.
Figure \ref{fig1}A shows the phases characterized by the active fraction $f_a$, with the absorbing--active (continuous) transition marked by the leftmost phase boundary defined by the critical density $\phi_c (\epsilon)$.  
Figure \ref{fig1}B shows the phases characterized by their global orientational order $|\langle\Psi_6\rangle| = |\langle(1/N_i)\sum_{j=1}^{N_i} e^{i6\theta_{ij}}\rangle_N|$, for nearest-neighbor bond angles $\theta_{ij}$. 
$|\langle\Psi_6\rangle|$ shows a second (discontinuous) phase boundary between active liquid and active crystal phases.
{Although difficult to see in Figure \ref{fig1}a, the discontinuous decrease in $|\langle\Psi_6\rangle|$ for increasing $\epsilon$ is accompanied by a discontinuous increase in the active fraction $f_a$, which is prominent as the melting boundary approaches the absorbing boundary but persists at larger densities (see Supplemental Figure S5).}
In contrast to one-dimensional BRO, where an ordered phase only exists at the critical point $\phi_c = 1$, crystalline order persists for all small-$\epsilon$ active states (see Supplemental Figure S8 for scaling of the melting transition in the high-density limit).
Thus, there are two phase transitions: the familiar absorbing-active transition and the active crystal-active disordered transition, which is unique to two dimensions. %The emergence of two separate active phases is unlike the behavior observed in 2D RO or 3D BRO, both of which have a single absorbing phase and a single disordered active phase.

\begin{figure}[b]%[hbt!]
   \centering
   \includegraphics[width=.5\textwidth]{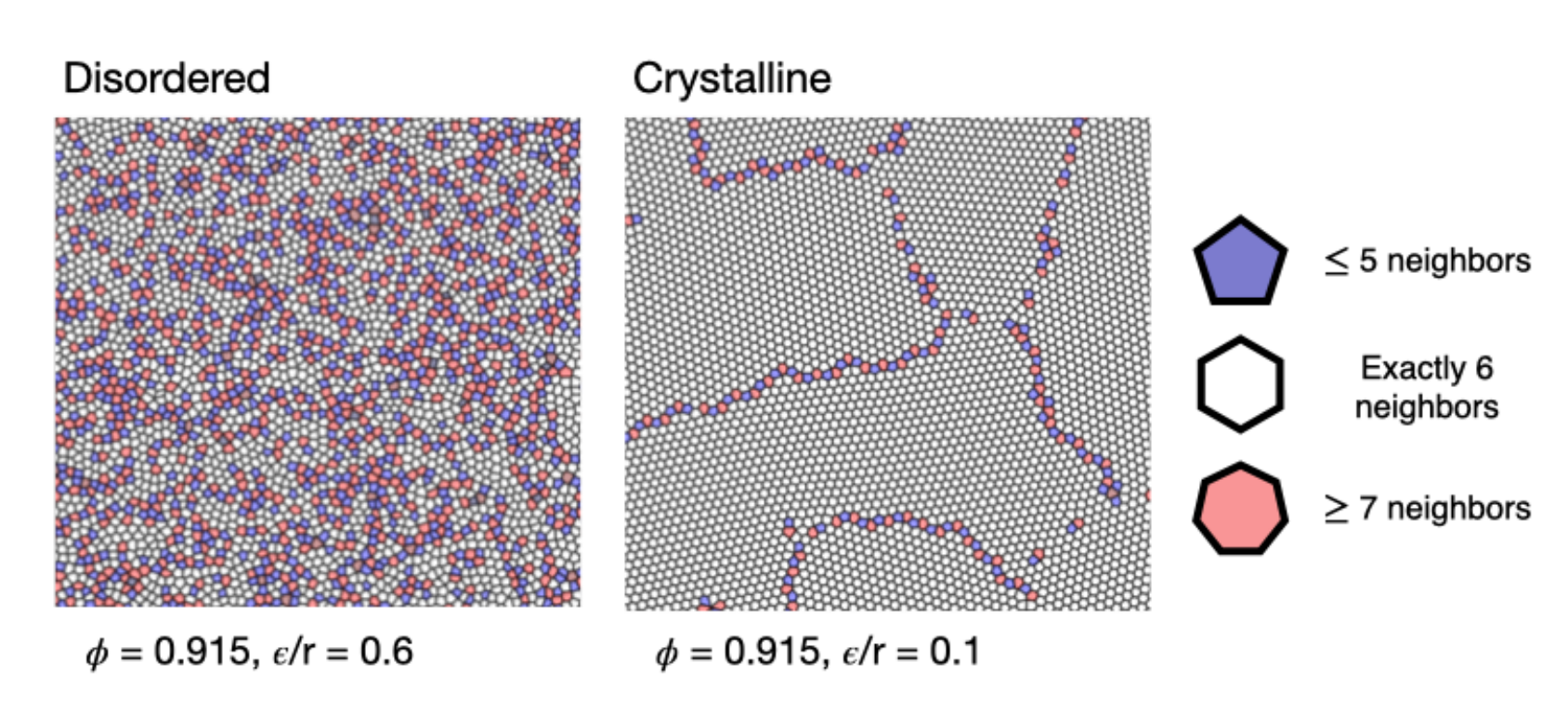}
    \caption{Two active configurations.  The left panel is in the active disordered phase, the right panel in the active crystalline phase.  Defects, particles with either fewer or more than 6 Voronoi neighbors, are indicated.  If we wait longer, the number of defects in the crystalline phase will decrease in a way reminiscent of Ostwald ripening.
    }
    \label{fig2}
\end{figure}

\begin{figure*}%[t]
    \centering
    \includegraphics[width=.9\textwidth]{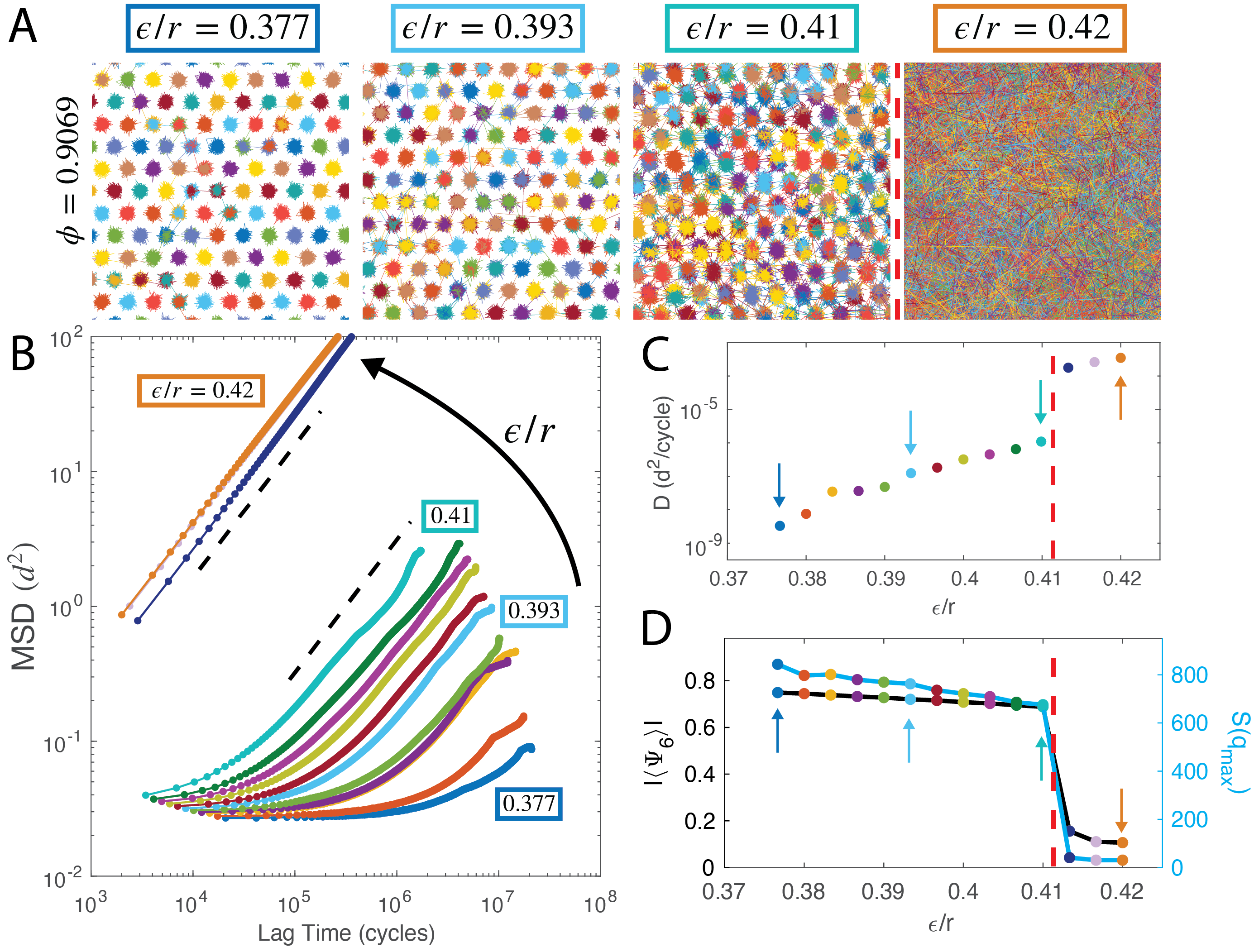}
    \caption{
    Active crystal dynamics.
    {\bf A} Trajectories of 2D BRO active defect-free crystals. Particles are largely confined to crystalline cages. However, crystal rearrangements occur when a large number of particles collectively hop lattice sites in a closed ring-like structure.
    Ring-like lattice diffusion occurs with increasing frequency as the melting boundary is approached $\epsilon_{melt} \approx 0.4125$.
    Above the melting point, long-range translational and orientational order are lost, and trajectories are isotropic and random.
    {\bf B} Mean-squared displacement is plotted for varied kick sizes crossing through the melting point.
    Liquid phases exhibit diffusive behavior (MSD $\sim t$) for $\epsilon/r \gtrsim \epsilon_{melt}$. Active crystalline phases exhibit long-time diffusive behavior and an intermediate caging plateau that increases with distance from the melting boundary. Colors correspond to the plotted points in {\bf C} and {\bf D}.
    {\bf C} The diffusion constant $D$, estimated from long-time dynamics, is plotted as a function of epsilon, showing a sharp, two-order-of-magnitude jump crossing through melting. Active crystals exhibit an exponentially decreasing diffusion coefficient away from the phase boundary. 
    {\bf D} The global orientational order $|\langle \Psi_6 \rangle|$ (black line) and global translational order $S(q_{max})$ (cyan line) both vanish simultaneously with increasing epsilon, indicating there is no hexatic phase. The sharp decrease in order is coincident with the jump in the diffusion coefficient.
    All simulations run with $N=1600$ particles.
    }
    \label{fig3}
\end{figure*}

%\textcolor{blue}{\it The next several paragraphs seem to be out of place - What are we trying to say?  Do we want to detail the different phases?  That would be ok, and here would be a bood place to do it, but we should say that at the outset.  As it stands, I would continue with the discussion on the phase transitions between the different phases on the next page, where I have indicated it in blue.}

Examples of active disordered and active crystalline configurations are shown in Figure \ref{fig2} at the same density $\phi = 0.915$ for large and small values of $\epsilon$, respectively.  
% There is a discontinuous jump in activity (though small) discussed previously
%It is important to remember that essentially all the particles in these snapshots are active at this density, that is, they overlap with their neighbors at least a little.  
Hexagonal crystal defects are indicated by color for particles with $n\neq6$ Voronoi neighbors.
% We don't discuss dislocations/disclinations in detail so is it necessary to define?
%, typically 5 or 7, where a 5-7 pair may be thought of as a dislocation, and an isolated 5 or 7 a positive or negative disclination, respectively.  
As seen in the right panel, the defects aggregate to form grain boundaries separating perfect hexagonal phases.  
Away from the absorbing boundary, grain boundaries coarsen with time, and the defect fraction decays as $t^{-1/3}$, reminiscent of Ostwald ripening~\cite{Voorhees1985}. 
Near the active-absorbing crystal transition, Ostwald ripening is retarded by critical slowing down dynamics (active fraction $f_a(t,\phi\rightarrow\phi_c)\sim t^{-\alpha}$) and the defect fraction decays like $t^{-1/3(1-\alpha)}$  (see Supplemental Figure S6).
In steady state, the defect fraction decreases sharply as $\epsilon$ decreases approximately exponentially across the melting boundary. 
%This functional form suggests that the active crystal phase contains a finite positive defect fraction for all $\epsilon < \epsilon_{melt}$ but contains no defects in the $\epsilon \rightarrow 0$ limit (see Supplemental Figure S6).
% Bring this back in the conclusion?
%It is intriguing to ask whether after a sufficiently long time all all the defects will disappear, leaving a perfect crystal.

We further observe this approach towards a defect-free hexagonal crystal by monitoring the evolution of the system after it is initialized deep in the polycrystalline region ($\epsilon / r =0.2, \phi =0.916$) with one side of the system disordered and the other side in a crystalline configuration (see Supplemental Figure S1), which shows the hexagonal lattice remaining defect--free while the disordered side of the system forms a polycrystalline structure. With time, grain boundaries in the polycrystalline region shrink as separate grains merge and grow, and growth of the hexagonal crystal extends into the polycrystalline region, suggesting that after sufficient time, the system will evolve to a defect-free crystal.

\begin{figure*}
    \centering
    \includegraphics[width=0.9\textwidth]{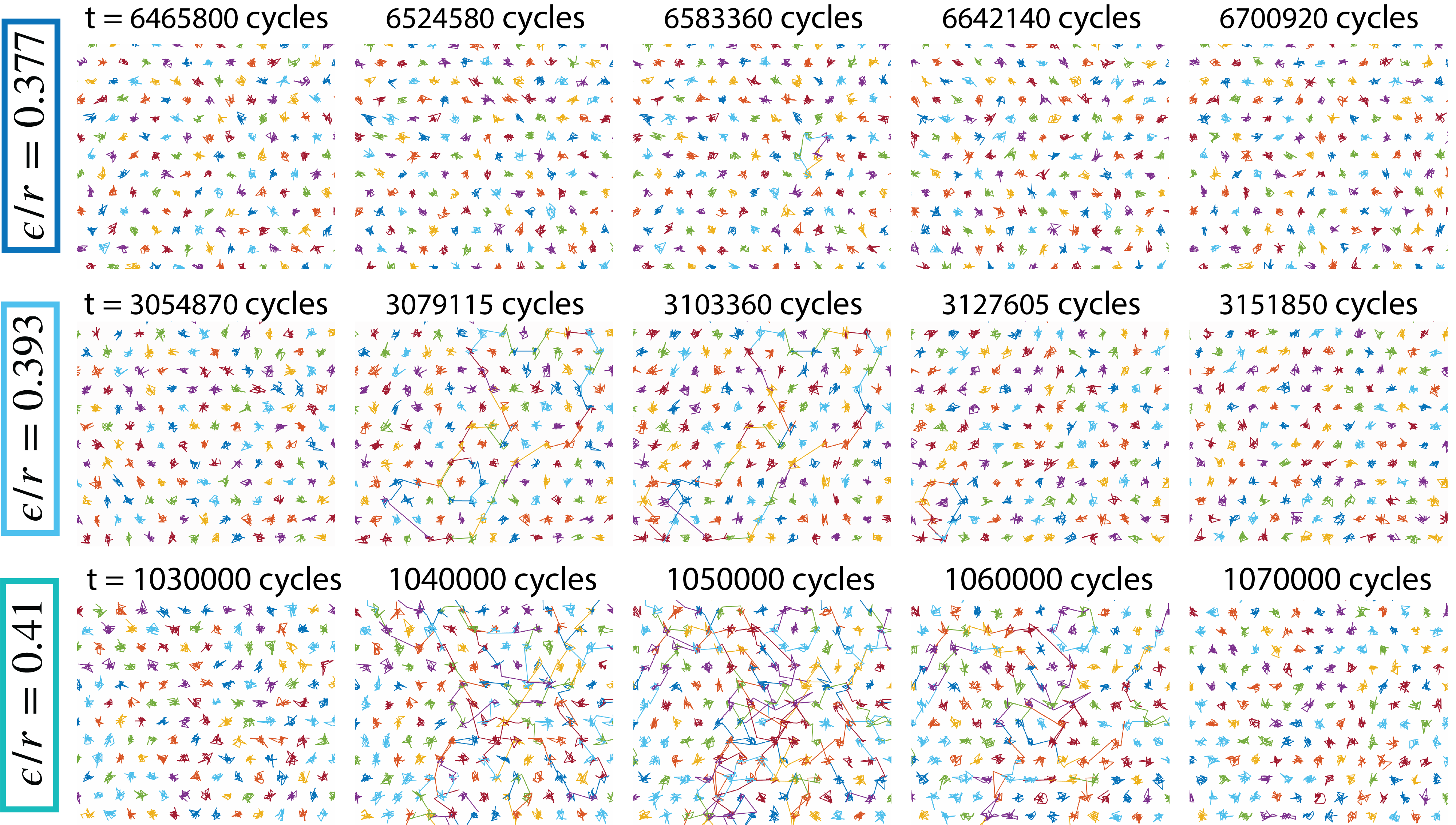}
    \caption{%\textcolor{red}{\bf Should this be horizontal rather than vertical?\\}
    Sequential trajectory images of a portion of the full system at $\phi = 0.9069$ for three kick sizes $\epsilon$ in the active crystal phase. 
    All $\epsilon$ values show the defect-free exchange of particles in a closed, ring shape.
    On approach to melting, loop structures grow in size and complexity, but the crystal phase remains stable. 
    Particles are labeled by identity.
    Trajectories are composed of 20 points recorded at time intervals of $\delta t= [2,4.8,11.8]\cdot 10^3$ cycles for $\epsilon/r = [0.41,0.393,0.377]$, respectively.
    }
    \label{fig3a}
\end{figure*}

% \begin{figure}
%     \centering
%     \includegraphics[width=0.45\textwidth]{Hopperfrac_stabilizerfrac_Fig.png}
%     \caption{{\bf A} The distribution of displacement magnitudes $|\Delta r|$ (in units of lattice spacing $l$) between frames $\delta t$ for varying kick size $\epsilon/r$.
%     In the crystal phase, displacements are primarily small, less than $0.9l$ (called `stabilizers'), with a small finite fraction of particles that hop between lattice sites (called `hoppers').
%     Above the melting point, the liquid phase displacement distribution is broad.
%     Distributions are colored to match main text Fig.~3.
%     Inset: Diagram shows six stabilizing particles surrounding a hopper. Arrows correspond to typical~$|\Delta r|$
%     {\bf B} The hopper (blue) and stabilizer (orange) fractions, plotted as a function of kick size, are mostly flat inside the crystal phase and then increase near the melting point, making a discontinuous jump when crossing it. Plotting the stabilizer fraction divided by six reveals that just below melting, six times the stabilizer fraction approaches the hopper fraction. Therefore, melting occurs when there is more than one hopper per six stabilizing particles.
%     }
%     \label{fig3a}
% \end{figure}

\begin{figure*}
    \centering
    \includegraphics[width=0.9\textwidth]{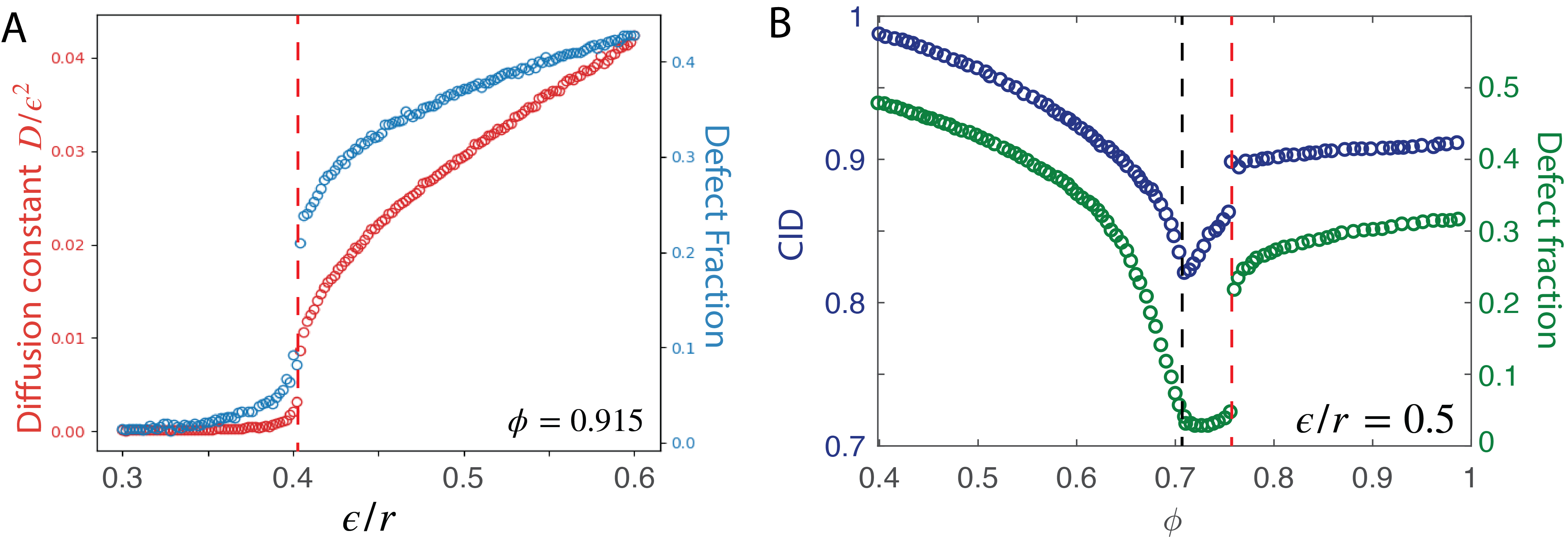}
    \caption{%\textcolor{red}{\bf Should this be horizontal rather than vertical?\\}
    {\bf A} The long-time diffusion constant (red) and the defect fraction (blue) are plotted as a function of kick size $\epsilon/r$ crossing the melting boundary at constant area fraction $\phi = 0.915$. 
    {\bf B} The computable information density (CID), blue points, and the defect fraction, green points, are plotted as a function of area fraction for constant $\epsilon/r$ where both second-order, absorbing-active crystal (black dashed line) and first-order active crystal-active liquid (red dashed line) phase boundaries are present. CID and defect fraction both decrease with increasing $\phi$. CID and defect fraction are continuous, but the slope changes sign crossing the absorbing-active boundary. 
    Both then show a discontinuous jump crossing the melting boundary.
    CID exhibits a stronger signal than the trend in defect density, indicating that structural correlations beyond the presence of defects also contribute to CID. 
    }
    \label{fig4}
\end{figure*}

%Taken together, the activity and global orientation order parameters (Figure \ref{fig1}), show that the phase behavior of 2D BRO does not simply consist of a single absorbing-active transition. 
%In addition to an activation transition, 
The competition between crystalline ordering (driven by small $\epsilon$ kicks) and disorder (promoted by large $\epsilon$) gives rise to a reentrant melting transition.  
This transition can take place both as a function of density, at intermediate kick sizes in the range $0.4 \lesssim \epsilon/r \lesssim 0.6$, and as a function of $\epsilon$, in the density range $0.7 \lesssim \phi \lesssim 0.9$. 
Increasing density first activates the system from a static disordered absorbing state to an active ordered crystalline state, but further increasing the density melts the system again into an active, disordered phase. 
Note that the active crystal phase continues to large $\phi$ as $\epsilon \rightarrow 0$.  In this large $\phi$ regime, the phase boundary goes as $\epsilon_\mathrm{melting} \sim \phi^{-1/2}$ .

The absorbing-active phase transition belongs to the Manna (conserved directed percolation) universality class, consistent with RO and BRO in other dimensions (see Supplementary Materials). 
We therefore focus on the new melting transition between two active phases. 
To better understand the nature of these active phases and the transition between them, we examine the motion of the individual particles.  It is more convenient to study this transition by changing $\epsilon$. 
In Figure \ref{fig3}, we study the individual particle trajectories and their mean squared displacement (MSD) as $\epsilon$ crosses the phase boundary, for $\phi_{hex} = \frac{\pi}{2\sqrt{3}} = 0.9069$.
There is nothing notable about this density in this range of $\epsilon$; the same behavior is seen for other densities along the liquid-crystal boundary (see supplemental figure S4).

Figure \ref{fig3}A shows the trajectories of individual particles leading up to and after crossing the melting transition. 
The loci of particle positions, colored by identity, are plotted for the last half of the simulation, where structures and dynamics have reached steady state.
Simulation runtimes are increased logarithmically (e.g. $t_{tot} = [2\cdot 10^7,8\cdot 10^6,3\cdot 10^6,2\cdot 10^6 ]$ for $\epsilon/r = [0.377,0.393,0.41,0.42]$) to account for slow crystal rearrangement dynamics deep into the active crystal phase.
As seen in the first three images of Figure \ref{fig3}A, at smaller kick sizes, the particles are active and primarily localized to lattice sites. 
However, particles occasionally collectively rearrange via a ring-like mechanism, where groups of particles jump lattice sites along a closed one-dimensional path (see Fig.~4 and Supplemental Videos 1-4).
While collective one-dimensional rearrangements have been previously observed in amorphous glassy systems~\cite{glotzer2000spatially}, this closed, loop exchange mechanism most closely resembles an instantaneous collective ring diffusion process proposed by Zener~\cite{zener1950ring}, which has rarely been observed~\cite{freysoldt2014first,pandey1986diffusion,montalenti2002spontaneous}. 
These rearrangements can be seen clearly in Supplemental movies of particle trajectories.
%As $\epsilon$ increases, rearrangements become more common. 
As $\epsilon$ increases, these rearrangements increase frequency, and the trajectories become more delocalized, until the melting transition takes place and the lattice is lost $\epsilon_{melt}/r \approx 0.4125$.
%, as seen in the $\epsilon/r = 0.42$ panel. 
%This bit is moved later
%This is the central characteristic of the active crystal phase: even when the particles are delocalized and hop from site to site, the lattice structure is maintained. 
%This effect violates the Lindemann criterion, a phenomenological metric that dictates that melting occurs when particles move on average a finite (~0.1 diameter) displacement from their average position~\cite{}. 

Lattice delocalization is also evident in the mean squared displacement (Figure \ref{fig3}B). 
We observe a long-time diffusive scaling ($\mathrm{MSD}\sim t$), allowing for a well-defined diffusion constant $D$ (Figure \ref{fig3}C). 
By plotting the diffusion constant across the melting transition, we observe a discontinuous jump at melting, indicating a first-order mobility transition. 
This is the central characteristic of the active crystal phase: even when the particles are delocalized and hop from site to site, the lattice structure is maintained. 
This effect violates the Lindemann criterion, a phenomenological metric that dictates that melting occurs when particles move on average a finite ($\approx 0.1$ diameter) displacement from their average position~\cite{lindemann1910ueber}. 

The transition between active crystal and isotropic active liquid occurs at a single point, in contrast to the two-step melting often observed in equilibrium. 
The global orientation order $|\langle\psi_6\rangle|$ and global translational order (i.e. the first Bragg peak of the structure factor $S(q_\mathrm{max})$), exhibit a simultaneous discontinuous jump at $\epsilon_c$, shown in Figure \ref{fig3}D (see Supplemental Figure S7 for full structure factors).
The first-order melting transition is also observed in the defect density, in Figure \ref{fig4}A, and the activity (Supplemental Figure S4).

Thus, the re-entrant disorder--order--disorder path consists of a second-order absorbing - active crystal transition followed by a first-order active crystal - active disordered transition.  
By taking a cut through the phase diagram at a constant, moderate kick size $\epsilon/r=0.5$ (Figure \ref{fig4}B), we show both the defect density and the Computable Information Density (CID) \cite{Martiniani2019}. 
CID is closely related to the Shannon entropy\cite{Shannon1948} and reduces to the thermodynamic entropy in equilibrium. 
The CID has been shown to exhibit singularities at non-equilibrium phase transitions\cite{Martiniani2019}, with first-order transitions showing a jump in the CID and second-order transitions showing a cusp. 
Both of these features are evident in the CID profile shown in Figure \ref{fig4}B as we pass from the absorbing phase to the active crystalline phase and then to the active disordered phase as density increases.

As the absorbing--active phase boundary is approached by increasing $\phi$, defects are greatly reduced, as is the CID, implying increased order, with the cusp in the CID signaling a second-order phase transition upon reaching the active crystalline phase. 
As the system crosses through the active crystalline phase, the defect density remains nearly constant. At the same time, the CID increases more sharply, suggesting that the placement of defects in the system is more random as density increases in the active crystalline phase. 
Finally, the melting transition from active crystalline to active disordered is marked by a discontinuous jump in both the CID and the defect density, again indicating a first-order transition and demonstrating a reentrance into a disordered phase after passing through the highly ordered active crystalline phase. 

In typical order-disorder transitions, there are two competing influences: one that promotes the formation of order, while the other opposes it. 
For systems whose particles interact via a potential, reducing the system energy promotes order, while the temperature randomizes. 
Perhaps the most surprising aspect of the order-disorder transition we have studied is that there is only one parameter at work, the kick size $\epsilon$.  
When $\epsilon$ is small, the system spontaneously self-organizes via small motions into the densest possible packing, a hexagonal crystal.  
This is true not only for densities near the hard disk packing limit $\phi_{hex}$, but for arbitrarily high densities and overlaps.  
It is this tendency to self-organize that mimics the role of lowering the energy by ordering, and it is the random small kicks that facilitate the self-organization.  
For large $\epsilon$, the randomness of the particles' displacement amplitudes prevents the particles from self-organizing into the dense packing.  
Thus, the random kicks both order and disorder the system. 

Possibly the most interesting aspect of these active crystals is that they remain crystalline even though the particles are not associated with a particular lattice site. Rather, particles cooperatively exchange lattice sites in closed rings. They diffuse through the lattice, spending most of their time near lattice sites but hopping frequently from one site to another.

\begin{acknowledgments}
We thank Yariv Kafri, Ludovic Berthier, and Lilith Lulovitch for useful discussions. 
This work was supported by the National Science Foundation Physics of Living Systems Grant No. 1504867 and U.S. Department of Energy, Office of Science, Office of Basic Energy Sciences under Award Number DE-SC-0020976 (A.G.) for simulations, the Center for Bio-Inspired Energy Sciences (CBES), an Energy Frontier Research Center funded by the U.S. DOE, Office of Sciences, Basic Energy Sciences under Award No. DE-SC0000989 and NSF CBET 1832260: GOALI (P.C.) for theory and calculations, and the MRSEC Program of the National Science Foundation under Grant No. DMR-1420073 (S.W.) for comparison simulations. D.L. acknowledges the support of the Israel Science Foundation (Grant No. 2083/23).
\end{acknowledgments}

%\normalem
%\bibliography{references}% Produces the bibliography via BibTeX.


\end{document}